# ADVANCED SPACE PROPULSION BASED ON VACUUM (SPACETIME METRIC) ENGINEERING

**HAROLD E. PUTHOFF**
*Institute for Advanced Studies at Austin, 11855 Research Blvd., Austin, Texas 78759, USA.*
Email: puthoff@earthtech.org

A theme that has come to the fore in advanced planning for long-range space exploration is the concept that empty space itself (the quantum vacuum, or spacetime metric) might be engineered so as to provide energy/thrust for future space vehicles. Although far-reaching, such a proposal is solidly grounded in modern physical theory, and therefore the possibility that matter/vacuum interactions might be engineered for space-flight applications is not *a priori* ruled out [1]. As examples, the current development of theoretical physics addresses such topics as warp drives, traversable wormholes and time machines that provide for such vacuum engineering possibilities [2-6]. We provide here from a broad perspective the physics and correlates/consequences of the engineering of the spacetime metric.

**Keywords:** Space propulsion, metric engineering, spacetime alteration, warp drives, wormholes, polarizable vacuum

## 1. INTRODUCTION

The concept of "engineering the vacuum" found its first expression in the physics literature when it was introduced by Nobelist T.D. Lee in his textbook *Particle Physics and Introduction to Field Theory* [7]. There he stated: "The experimental method to alter the properties of the vacuum may be called vacuum engineering.... If indeed we are able to alter the vacuum, then we may encounter new phenomena, totally unexpected." This legitimization of the vacuum engineering concept was based on the recognition that the vacuum is characterized by parameters and structure that leave no doubt that it constitutes an energetic and structured medium in its own right. Foremost among these are that (1) within the context of quantum theory the vacuum is the seat of energetic particle and field fluctuations, and (2) within the context of general relativity the vacuum is the seat of a spacetime structure (metric) that encodes the distribution of matter and energy. Indeed, on the flyleaf of a book of essays by Einstein and others on the properties of the vacuum we find the statement "The vacuum is fast emerging as *the* central structure of modern physics" [8]. Perhaps the most definitive statement acknowledging the central role of the vacuum in modern physics is provided by 2004 Nobel Prize winner Frank Wilczek in his recent book *The Lightness of Being: Mass, Ether and the Unification of Forces* [9]:

> "What is space? An empty stage where the physical world of matter acts out its drama? An equal participant that both provides background and has a life of its own? Or the primary reality of which matter is a secondary manifestation? Views on this question have evolved, and several times have changed radically, over the history of science. Today the third view is triumphant."

Given the known characteristics of the vacuum, one might reasonably inquire as to why it is not immediately obvious how to catalyze robust interactions of the type sought for space-flight applications. To begin, in the case of quantum vacuum processes there are uncertainties that remain to be clarified regarding global thermodynamic and energy constraints. Furthermore, it is likely that energetic components of potential utility involve very small-wavelength, high-frequency field structures and thus resist facile engineering solutions. With regard to perturbation of the spacetime metric, the required energy densities predicted by present theory exceed by many orders of magnitude values achievable with existing engineering techniques. Nonetheless, one can examine the possibilities and implications under the expectation that as science and its attendant derivative technologies mature, felicitous means may yet be found that permit the exploitation of the enormous, as-yet-untapped potential of engineering so-called "empty space," the vacuum.

In Section 2 the underlying mathematical platform for investigating spacetime structure, the metric tensor approach, is introduced. Section 3 provides an outline of the attendant physical effects that derive from alterations in the spacetime structure, and Section 4 catalogs these effects as they would be exhibited in the presence of advanced aerospace craft technologies based on spacetime modification.

## 2. SPACETIME MODIFICATION – METRIC TENSOR APPROACH

Despite the daunting energy requirements to restructure the spacetime metric to a significant degree, the forms that such restructuring would take to be useful for space-flight applications can be investigated, and their corollary attributes and consequences determined - a "Blue Sky," general-relativity-for-engineers approach, as it were. From such a study the signatures that would accompany such advanced-technology craft can be outlined, and possible effects of the technology with regard to spacetime effects that include such phenomena as the distortion of space and time can be cataloged. This would include, among other consequences, cataloging effects that might be potentially harmful to human physiology.

The appropriate mathematical evaluation tool is use of the *metric tensor* that describes the measurement of spacetime intervals. Such an approach, well-known from studies in GR (general relativity) has the advantage of being model-independent, i.e., does not depend on knowledge of the specific mechanisms or dynamics that result in spacetime alterations, only that a technology exists that can control and manipulate (i.e., engineer) the spacetime metric to advantage. Before discussing the predicted characteristics of such engineered spacetimes a brief mathematical digression is in order for those interested in the mathematical structure behind the discussion to follow.

As a brief introduction, the expression for the 4-dimensional line element $ds^2$ in terms of the metric tensor $g_{\mu\nu}$ is given by

$$ds^2 = g_{\mu\nu} dx^\mu dx^\nu \quad (1)$$

where summation over repeated indices is assumed unless otherwise indicated. In ordinary Minkowski flat spacetime a (4-dimensional) infinitesimal interval $ds$ is given by the expression (in Cartesian coordinates)

$$ds^2 = c^2 dt^2 - (dx^2 + dy^2 + dz^2) \quad (2)$$

where we make the identification $dx^0 = cdt$, $dx^1 = dx$, $dx^2 = dy$, $dx^3 = dz$, with metric tensor coefficients $g_{00} = 1$, $g_{11} = g_{22} = g_{33} = -1$, $g_{\mu\nu} = 0$ for $\mu \neq \nu$.

For spherical coordinates in ordinary Minkowski flat spacetime

$$ds^2 = c^2 dt^2 - dr^2 - r^2 d\theta^2 - r^2 \sin^2\theta d\varphi^2 \quad (3)$$

where $dx^0 = cdt$, $dx^1 = dr$, $dx^2 = d\theta$, $dx^3 = d\varphi$, with metric tensor coefficients $g_{00} = 1$, $g_{11} = -1$, $g_{22} = -r^2$, $g_{33} = -r^2 \sin^2\theta$, $g_{\mu\nu} = 0$ for $\mu \neq \nu$.

As an example of spacetime alteration, in a spacetime altered by the presence of a spherical mass distribution $m$ at the origin (Schwarzschild-type solution) the above can be transformed into [10]

$$ds^2 = \left(\frac{1 - Gm/rc^2}{1 + Gm/rc^2}\right) c^2 dt^2 - \left(\frac{1 - Gm/rc^2}{1 + Gm/rc^2}\right)^{-1} dr^2 \\ - \left(1 + Gm/rc^2\right)^2 r^2 \left(d\theta^2 + \sin^2\theta d\varphi^2\right) \quad (4)$$

with the metric tensor coefficients $g_{\mu\nu}$ modifying the Minkowski flat-spacetime intervals $dt$, $dr$, etc., accordingly.

As another example of spacetime alteration, in a spacetime altered by the presence of a *charged* spherical mass distribution $(Q, m)$ at the origin (Reissner-Nordstrom-type solution) the above can be transformed into [11]

$$ds^2 = \left(\frac{1 - Gm/rc^2}{1 + Gm/rc^2} + \frac{Q^2 G/4\pi\varepsilon_0 c^4}{r^2 \left(1 + Gm/rc^2\right)^2}\right) c^2 dt^2 \\ - \left(\frac{1 - Gm/rc^2}{1 + Gm/rc^2} + \frac{Q^2 G/4\pi\varepsilon_0 c^4}{r^2 \left(1 + Gm/rc^2\right)^2}\right)^{-1} dr^2 \\ - \left(1 + Gm/rc^2\right)^2 r^2 \left(d\theta^2 + \sin^2\theta d\varphi^2\right) \quad (5)$$

with the metric tensor coefficients $g_{\mu\nu}$ again changed accordingly. In passing, one can note that the effect on the metric due to charge $Q$ differs in sign from that due to mass m, leading to what in the literature has been referred to as *electrogravitic repulsion* [12].

Similar relatively simple solutions exist for a spinning mass (Kerr solution), and for a spinning electrically charged mass (Kerr-Newman solution). In the general case, appropriate solutions for the metric tensor can be generated for arbitrarily-engineered spacetimes, characterized by an appropriate set of spacetime variables $dx^\mu$ and metric tensor coefficients $g_{\mu\nu}$. Of significance now is to identify the associated physical effects and to develop a Table of such effects for quick reference.

The first step is to simply catalog metric effects, i.e., physical effects associated with alteration of spacetime variables, and save for Section 4 the significance of such effects within the context of advanced aerospace craft technologies.

### 3. PHYSICAL EFFECTS AS A FUNCTION OF METRIC TENSOR COEFFICIENTS

In undistorted spacetime, measurements with physical rods and clocks yield spatial intervals $dx^\mu$ and time intervals $dt$, defined in a flat Minkowski spacetime, the spacetime of common experience. In spacetime-altered regions, we can still choose $dx^\mu$ and $dt$ as *natural* coordinate intervals to represent a coordinate map, but now local measurements with *physical* rods and clocks yield spatial intervals

$$\sqrt{-g_{\mu\nu}} dx^\mu$$

and time intervals

$$\sqrt{g_{00}} dt$$

so-called *proper* coordinate intervals. From these relationships a Table can be generated of associated physical effects to be expected in spacetime regions altered by either natural or advanced technological means. Given that, as seen from an unaltered region, alteration of spatial and temporal intervals in a spacetime-altered region result in an altered velocity of light, from an engineering viewpoint such alterations can in essence be understood in terms of a variable refractive index of the vacuum (see Section 3.4 below) that affects all measurement.

#### 3.1 Time Interval, Frequency, Energy

The case where

$$\sqrt{g_{00}} < 1$$

is considered first, typical for an altered spacetime metric in the vicinity of, say, a stellar mass – see leading term in Eq. (4). Local measurements with physical clocks within the altered spacetime region yield a time interval

$$\sqrt{g_{00}} dt < dt$$

thus an interval of time $dt$ between two events located in an undistorted spacetime region remote from the mass (i.e., an observer at infinity) – say, ten seconds - would be judged by local (proper) measurement from *within* the altered spacetime region to occur in a lesser time interval,

$$\sqrt{g_{00}} dt < dt$$

- say, 5 seconds. From this it can be rightly inferred that, relatively speaking, clocks (including atomic processes, etc.) within the altered spacetime run slower. Given this result, a physical process (e.g., interval between clock ticks, atomic emissions, etc.) that takes a time $\Delta t$ in unaltered spacetime slows to

$$\Delta t \to \Delta t / \sqrt{g_{00}}$$

when occurring within the altered spacetime. Conversely, under conditions (e.g., metric engineering) for which

$$\sqrt{g_{00}} > 1$$

processes within the spacetime-altered region are, relatively speaking, sped up. Thus we have our first entry for a Table of physical effects (see Table 1).

Given that frequency measurements are the reciprocal of time duration measurements, the associated expression for frequency $\omega$ is given by

$$\omega \to \omega \sqrt{g_{00}}$$

the second entry in Table 1. This accounts, e.g., for the redshifting of atomic emissions from dense masses where

$$\sqrt{g_{00}} < 1$$

Conversely, under conditions for which

$$\sqrt{g_{00}} > 1$$

blueshifting of emissions would occur. In addition, given that quanta of energy are given by $E = \hbar \omega$, energy scales with

$$\sqrt{g_{00}}$$

as does frequency,

$$E \to E \sqrt{g_{00}}$$

the third entry in Table 1. Depending on the value of

$$\sqrt{g_{00}}$$

in the spacetime-altered region, energy states may be raised or lowered relative to an unaltered spacetime region.

### 3.2 Spatial Interval

Again, by considering the case typical for an altered spacetime metric in the vicinity of, say, a stellar mass,

$$\sqrt{-g_{11}} > 1$$

for the radial dimension $x^1 = r$ – see second term in Eq. (4). Therefore, local measurements with physical rulers within the altered spacetime yield a spatial interval

$$\sqrt{-g_{11}} \, dr > dr$$

thus a spatial interval $dr$ between two locations in an undistorted spacetime (say, remote from the mass) would be judged by local (proper) measurement from within the altered spacetime to be greater,

$$\sqrt{-g_{11}} \, dr > dr$$

From this it can be rightly inferred that, relatively speaking, rulers (including atomic spacings, etc.) within the altered spacetime are shrunken relative to their values in unaltered spacetime. Given this result, a physical object (e.g., atomic orbit) that possesses a measure $\Delta r$ in unaltered spacetime shrinks to

$$\Delta r \to \Delta r / \sqrt{-g_{11}}$$

when placed within the altered spacetime. Conversely, under conditions for which

**TABLE 1:** *Metric Effects on Physical Processes in an Altered Spacetime as Interpreted by a Remote (Unaltered Spacetime) Observer.*

| Variable | Typical Stellar Mass $(g_{00} < 1, \ |g_{11}| > 1)$ | Spacetime-Engineered Metric $(g_{00} > 1, \ |g_{11}| < 1)$ |
|---|---|---|
| **Time Interval** $\Delta t \to \Delta t / \sqrt{g_{00}}$ | processes (e.g., clocks) run slower | processes (e.g., clocks) run faster |
| **Frequency** $\omega \to \omega \sqrt{g_{00}}$ | red shift toward lower frequencies | blueshift toward higher frequencies |
| **Energy** $E \to E \sqrt{g_{00}}$ | energy states lowered | energy states raised |
| **Spatial measure** $\Delta r \to \Delta r / \sqrt{-g_{11}}$ | objects (e.g., rulers) shrink | objects (e.g., rulers) expand |
| **Velocity of light** $v_L = c \to c \sqrt{g_{00} / -g_{11}}$ | effective $v_L < c$ | effective $v_L > c$ |
| **Mass** $m$ $m = E/c^2$ $\to \left( -g_{11} / \sqrt{g_{00}} \right) m$ | effective mass increases | effective mass decreases |
| **Gravitational "force"** $f(g_{00}, g_{11})$ | "gravitational" | "antigravitational" |

$$\sqrt{-g_{11}} < 1$$

objects would expand - thus the fourth entry in Table 1 of physical effects.

### 3.3 Velocity of Light in Spacetime-Altered Regions

Interior to a spacetime region altered by, say, a dense mass (e.g., a black hole), the locally-measured velocity of light *c* in, say, the $x^1 = r$ direction, is given by the ratio of locally-measured (proper) distance/time intervals for a propagating light signal [13]

$$v_L^i = \frac{\sqrt{-g_{11}}\,dr}{\sqrt{g_{00}}\,dt} = c \qquad (6)$$

From a viewpoint *exterior* to the region, however, from the above the remotely-observed *coordinate* ratio measurement yields a different value

$$v_L^e = \frac{dr}{dt} = \sqrt{\frac{g_{00}}{-g_{11}}}\,c \qquad (7)$$

Therefore, although a local measurement with physical rods and clocks yields *c*, an observer in an exterior reference frame remote from the mass speaks of light "slowing down" on a radial approach to the mass due to the ratio

$$\sqrt{g_{00}/-g_{11}} < 1$$

Conversely, under (metric engineering) conditions for which

$$\sqrt{g_{00}/-g_{11}} > 1$$

the velocity of light - and advanced-technology craft velocities that obey similar formulae – would appear superluminal in the exterior frame. This yields the fifth entry in Table 1 of physical effects.

### 3.4 Refractive Index Modeling

Given that velocity-of-light effects in a spacetime-altered region, as viewed from an external frame, are governed by Eq. (7), it is seen that the effect of spacetime alteration on light propagation can be expressed in terms of an optical refractive index n, defined by

$$v_L^e = \frac{c}{n}, \qquad n = \sqrt{\frac{-g_{11}}{g_{00}}} \qquad (8)$$

where *n* is an effective refractive index of the (spacetime-altered) vacuum. This widely-known result has resulted in the development of refractive-index models for GR [14-17] that have found application in problems such as gravitational lensing [18]. The estimated electric or magnetic field strengths required to generate a given refractive index change are given by standard GR theory (the Levi-Civita Effect) and can be found in [19].

In engineering terms, the velocity of light *c* is given by the expression

$$c = 1/\sqrt{\mu_0 \varepsilon_0}$$

where $\mu_0$ and $\varepsilon_0$ are the magnetic permeability and dielectric permittivity of undistorted vacuum space ($\mu_0 = 4\pi \times 10^{-7}$ H/m and $\varepsilon_0 = 8.854 \times 10^{-12}$ F/m). The generation of an effective refractive index

$$n = \sqrt{-g_{11}/g_{00}} \neq 1$$

by technological means can from an engineering viewpoint be interpreted as manipulation of the vacuum parameters $\mu_0$ and $\varepsilon_0$. In GR theory such variations in $\mu_0$, $\varepsilon_0$, and hence the velocity of light, *c*, can be treated in terms of a "*THεμ*" formalism used in comparative studies of gravitational theories [20].

As discussed in Section 4.4 below, a number of striking effects can be anticipated in certain engineered spacetime regions.

### 3.5 Effective Mass in Spacetime-Altered Regions

In a spacetime-altered region $E = mc^2$ still holds in terms of local ("proper coordinate") measurements, but now energy *E* and the velocity of light *c* take on altered values as observed from an exterior (undistorted) spacetime region. Reference to the definitions for *E* and *c* in Table 1 permits one to define an effective mass as seen from the exterior undistorted region as therefore taking on the value

$$m \to m(-g_{11})/\sqrt{g_{00}}$$

providing a sixth entry for Table 1. Depending on the values of $g_{00}$ and $g_{11}$ the effective mass may be seen from the viewpoint of an observer in an undistorted spacetime region to have either increased or decreased.

### 3.6 Gravity/Antigravity "Forces"

Strictly speaking, in the GR point of view there are no gravitational "forces," but rather (in the words of GR theorist John Wheeler) "Matter tells space how to curve, and space tells matter how to move [21]." As a result, Newton's law of gravitational attraction to a central mass is therefore interpreted in terms of the spacetime structure as expressed in terms of the metric tensor coefficients, in this case as expressed in Eq. (4) above. Therefore, in terms of the metric coefficients gravitational attraction in this case derives from the condition that

$$g_{00} < 1, |g_{11}| > 1.$$

As to the possibility for generating "antigravitational forces," it was noted that in equation (5) inclusion of the effects of charge led to metric tensor contributions counter to the effects of mass, i.e., to *electrogravitic repulsion*. This reveals that conditions under which, say, the signs of the coefficients $g_{00}$ and $g_{11}$ could be reversed would be considered (loosely) as antigravitational in nature. A seventh entry in Table 1 represents these features of metric significance.

### 4. SIGNIFICANCE OF PHYSICAL EFFECTS APPLICABLE TO ADVANCED AEROSPACE CRAFT TECHNOLOGIES AS A FUNCTION OF METRIC TENSOR COEFFICIENTS

As discussed in Section 3, metric tensor coefficients define the relationship between locally- and remotely-observed (i.e., spacetime-altered and unaltered) variables of interest as listed in Table 1, and in the process define corollary physical effects. Table 1 thereby constitutes a useful reference for interpreting the physical significance of the effects of the alteration of spacetime variables. The expressions listed indicate specific spacetime alteration effects, whether due to natural causes (e.g., the presence of a planetary or stellar mass), or as a result

*Harlod E. Puthoff*

of *metric engineering* by advanced technological means as might be anticipated in the development and deployment of advanced aerospace craft operating on the principles of vacuum engineering.

### 4.1 Time Alteration

With regard to the first entry in Table 1 (time interval), in a spacetime-altered region time intervals are seen by a remote (unaltered spacetime) observer to vary as

$$1/\sqrt{g_{00}}$$

relative to the remote observer. Near a dense mass,

$$\sqrt{g_{00}} < 1$$

for example, and therefore time intervals are seen as lengthening and processes to run slower (and, in the case of a black hole, to stop altogether). One consequence of this is redshift of emission lines. Should such a time-slowed condition be engineered in an advanced aerospace application, an individual having spent time within such a temporally-modified field would, when returned to the normal environment, find that more time had passed than could be experientially accounted for. If uninformed about the metric-engineered characteristics of the environment from which he emerged, the individual might be inclined to interpret the experience in terms of a "missing time" *Rip van Winkle* effect, as it were.

Conversely, for an engineered spacetime associated with an advanced aerospace craft in which

$$\sqrt{g_{00}} > 1$$

time flow within the altered spacetime region would appear sped up to an external observer, while to an internal observer external time flow would appear to be in slow motion. In this scenario close approach to such a craft could leave one with the impression of, say, a 20 minute time interval (corroborated by the observer's watches), whereas only a few minutes would have passed in "real" or "normal" or "exterior" time. A corollary would be that within the spacetime-altered region normal environmental sounds from outside the region might cease to be registered, since external sounds could under these conditions redshift below the auditory range.

An additional implication of time speedup within the frame of such an exotic craft technology is that its flight path that might seem precipitous from an external viewpoint (e.g., sudden acceleration or deceleration) would be experienced as much less so by the craft's occupants. From the occupants' viewpoint, observing the external environment to be in relative slow motion, it would not be surprising to consider that one's relatively modest changes in motion would appear abrupt to an external observer.

Based on the second entry in Table 1 (frequency), yet another implication of an accelerated time-frame due to craft-associated metric engineering that leads to

$$\sqrt{g_{00}} > 1$$

frequencies associated with the craft would for a remote observer appear to be blueshifted. Corollary to observation of such a craft is the possibility that there would be a brightening of luminosity due to the heat spectrum blueshifting up into the visible portion of the spectrum (see Fig. 1). Additionally, close approach to such a craft could lead to possible harmful effects from ultraviolet and soft-X-ray generation due to blueshifting of the visible portion of the spectrum to higher frequencies.

With regard to the third entry in Table 1 (energy), in a spacetime-altered region energy scales as

$$\sqrt{g_{00}}$$

relative to a remote observer in an undistorted spacetime. In the vicinity of a dense mass where

$$\sqrt{g_{00}} < 1$$

the consequent reduction of energy bonds correlates with observed redshifts of emission. For engineered spacetimes associated with advanced craft technology in which

$$\sqrt{g_{00}} > 1$$

(accelerated time-frame case), a craft's material properties would appear "hardened" relative to the environment due to the increased binding energies of atoms in its material structure. Such a craft could, say, impact water at high velocities without apparent deleterious effects. A corollary is that the potential radiation exposure effects mentioned above would not be hazardous to craft occupants since for those totally within the field of influence the biological chemical bonds would be similarly hardened. Finally, an additional side effect potentially associated with exposure to the accelerated time-frame field would be accelerated aging of, say, plants in the area of a landed craft, and thus observation of the latter could act as a marker indicating the presence of such a field.

### 4.2 Spatial Alteration

The fourth entry in Table 1 (spatial measure) indicates the size of an object within an altered spacetime region as seen by a remote observer. The size of, say, a spherical object is seen to have its radial dimension, r, scale as

$$1/\sqrt{-g_{11}}$$

In the vicinity of a dense mass

$$\sqrt{-g_{11}} > 1$$

in which case an object within the altered spacetime region appears to a remote observer to have shrunk. As a corollary,

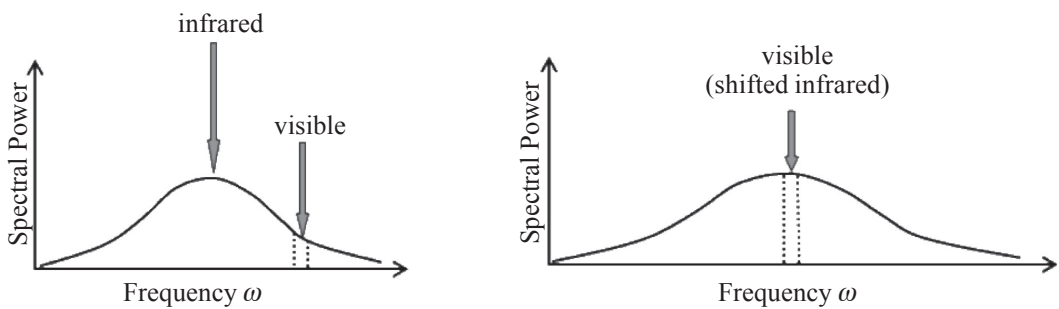

**Fig. 1 Blueshifting of infrared heat power spectrum.**

metric engineering associated with an exotic craft to produce this effect could in principle result in a large craft with spacious interior appearing to an external observer to be relatively small. Additional dimensional aspects such as potential dimensional changes are discussed in Section 4.4, "Refractive Index" Effects.

### 4.3 Velocity of Light/Craft in Spacetime-Altered Regions

Interior to a *spacetime-altered* region the locally-measured velocity of light, $v_L^i = c$, is given by the ratio of (locally-measured) distance/time intervals for a propagating light signal – Eq. (6) above. From a viewpoint exterior to the region, however, the observed coordinate ratio measurement can yield a different value $v_L^e$ greater or less than c as given by the 5th entry in Table 1. As an example of a measurement less than c, one speaks of light "slowing down" as a light signal approaches a dense mass (e.g., a black hole.) In an engineered spacetime in which,

$$g_{00} > 1, |g_{11}| < 1$$

however, the effective velocity of light $v_L^e$ as measured by an external observer can be $> c$.

Further, velocities in general in different coordinate systems scale as does the velocity of light, i.e.,

$$v \to \sqrt{g_{00}/-g_{11}}\, v$$

For exotic spacecraft an engineered spacetime metric - generated by and carried along with the craft - can in principle establish a condition in which the motion of a craft approaching the velocity of light in its own frame would be observed from an exterior frame to exceed light speed, i.e., exhibit motion at superluminal speed. This opens up the possibility of transport at superluminal velocities (as measured by an external observer) without violation of the velocity-of-light constraint within the spacetime-altered region, a feature attractive for interstellar travel. This is the basis for discussion of *warp drives* and *wormholes* in the GR literature [2-6]. Although present technological facility is far from mature enough to support the development of warp drive and wormhole technologies [22], the possibility of such technologies being developed in the future cannot be ruled out. In other words, effective transport at speeds exceeding the conventional speed of light could occur in principle, and therefore the possibility of reduced-time interstellar travel is not fundamentally ruled out by physical principles.

### 4.4 Refractive Index Effects

When one considers metric-engineered spacetimes associated with exotic propulsion, a number of corollary side effects associated with refractive index changes of the vacuum structure emerge as possibilities. Expected effects would mimic known refractive index effects in general, and can therefore be determined from known phenomena. Indistinct boundary definition associated with "waviness" as observed with heat waves off a desert floor is one example. As another, a light beam may bend (as in the GR example of the bending of starlight as it grazes the sun – see Fig. 2) or even terminate in mid-space. Such an observation would exhibit features that under ordinary circumstances would be associated with a high-refractive-index optical fiber in normal space (well-defined boundaries, light trapped within, bending or termination in mid-space). Additional observations might include apparent changes in size or shape (changes in lensing magnification parameters). Yet another possibility is the sudden "cloaking" or "blinking out," which would at least be consistent with strong gravitational lensing effects that bend a background view around a craft, though other technical options involving, e.g., the use of metamaterials, exist as well [23].

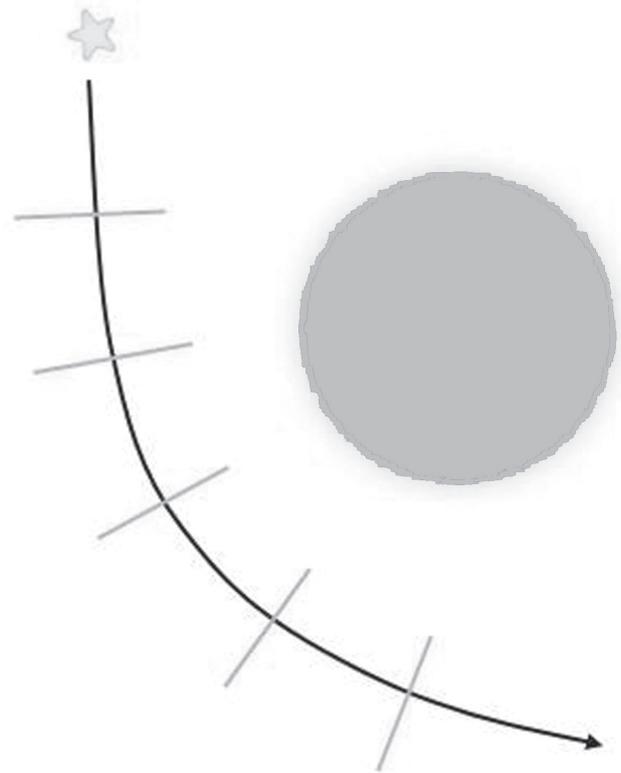

**Fig. 2** Light-bending in a spacetime-altered region.

### 4.5 Effective Mass in Spacetime-Altered Regions

As noted in Section 3.5, spacetime alteration of energy and light-speed measures leads to an associated alteration in the effective mass of an object in a spacetime-altered region as viewed from an external (unaltered) region. Of special interest is the case in which the effective mass is decreased by application of spacetime metric engineering principles as might be expected in the case of metric engineering for spaceflight applications (reference last column in Table 1). Effective reduction of inertial mass as viewed in our frame of reference would appear to mitigate against untoward effects on craft occupants associated with abrupt changes in movement. (The physical principles involved can alternatively be understood in terms of associated coordinate transformation properties as discussed in Section 4.1.) In any case, changes in effective mass associated with engineering of the spacetime metric in a craft's environs can lead to properties advantageous for spaceflight applications.

### 4.6 Gravity/Antigravity/Propulsion Effects

In the GR ansatz gravitational-type forces derive from the spacetime metric, whether determined by natural sources (e.g., planetary or stellar masses) or by advanced metric engineering. Fortunately for consideration of this topic, discussion can be carried out solely based on the form of the metric, independent of the specific mechanisms or dynamics that determine the metric. As one exemplar consider Alcubierre's formulation of a "warp drive," a spacetime metric solution of Einstein's GR

*Harlod E. Puthoff*

field equation [2, 22]. Alcubierre derived a spacetime metric motivated by cosmological inflation that would allow arbitrarily short travel times between two distant points in space. The behavior of the warp drive metric provides for the simultaneous expansion of space behind the spacecraft and a corresponding contraction of space in front of the spacecraft (see Fig. 3). The warp drive spacecraft would thus appear to be "surfing on a wave" of spacetime geometry. By appropriate structuring of the metric the spacecraft can be made to exhibit an arbitrarily large apparent faster-than-light speed as viewed by external observers without violating the local speed-of-light constraint within the spacetime-altered region. Furthermore, the Alcubierre solution showed that the proper (experienced) acceleration along the spaceship's path would be zero, and that the spaceship would suffer no time dilation, highly desirable features for interstellar travel. In order to implement a warp drive, one would have to construct a "warp bubble" that surrounded the spacecraft by generating a thin-shell or surface layer of exotic matter, i.e., a quantum field having negative energy and/or negative pressure. Although the technical requirements for such are unlikely to be met in the foreseeable future [22], nonetheless the exercise serves as a good example for showcasing attributes associated with manipulation of the spacetime metric at will.

The entire discussion of the possibility of generating a spacetime structure like that of the Alcubierre warp drive is based simply on assuming the form of a metric (i.e., $g_{\mu\nu}$) that exhibits desired characteristics. In like manner, arbitrary spacetime metrics to provide gravity/antigravity/propulsion characteristics can in principle be postulated. What is required for implementation is to determine appropriate sources for their generation, a requirement that must be met before advanced spaceship technology based on vacuum engineering can be realized in practice. The difficulties and challenges as well as the options for meeting such requirements can be found in the relevant literature [22].

## 5. DISCUSSION

In this paper the possibility has been considered that future developments with regard to advanced aerospace technologies could trend in the direction of manipulating the underlying spacetime structure of the vacuum of space itself by processes that can be called vacuum engineering or metric engineering. Far from being simply a fanciful concept, a significant literature exists in peer-reviewed, Tier 1 physics publications in which the topic is explored in detail. (See Ref. [1] for a comprehensive introduction to the subject with contributions from lead scientists from around the globe.)

The analysis presented herein, a form of general relativity (GR) for engineers so to speak, takes advantage of the fact that in GR a minimal-assumption, metric tensor approach can be used that is model-independent - that is, does not depend on knowledge of the specific mechanisms or dynamics that result in spacetime alterations, only assumes that a technology exists that can control and manipulate (i.e., engineer) the spacetime variables to advantage. Such an approach requires only that the hypothesized spacetime alterations result in effects consonant with our presently known GR physics principles.

In the metric engineering approach application of the principles gives precise predictions as to what can be expected as spatial and temporal variables are altered from their usual (i.e., flat space) structure. Signatures of the predicted contractions and expansions of space, slowdown and speedup of time, alteration of effective mass, speed of light and associated consequences, both as occur in natural phenomena in nature and with regard to spacetimes specifically engineered for advanced aerospace applications, are succinctly summarized in Table 1 included in the text.

Of particular interest with regard to innovative forms of advanced aerospace craft are the features tabulated in the right-hand column of the Table, features that presumably describe an ideal craft for interstellar travel: an ability to travel at superluminal speeds relative to the reference frame of background space, energy bonds of materials strengthened (i.e., hardened) relative to the background environment, a decrease in effective mass *vis à vis* the environment, an accelerated time frame that would permit rapid trajectory changes relative to the background rest frame without undue internal stress, and the generation of gravity-like forces of arbitrary geometry - all on the basis of restructuring the vacuum spacetime variables [24]. As *avant garde* as such features appear to be, they are totally in conformance with the principles of general relativity as presently understood. What remains as a challenge is to develop insight into the technological designs by which such vacuum restructuring can be generated on the scale required to implement the necessary spacetime modifications.

Despite the challenges, sample calculations as presented herein indicate the direction of potentially useful trends deriv-able on the basis of the application of GR principles as embodied in a metric engineering approach, with the results constrained only by what is achievable practically in an engineering sense. The latter is, however, a daunting constraint. At this point in the consideration of such nascent concepts, given our present level of technological evolution, it is premature to even guess at an optimum strategy, let alone attempt to form a critical path for the engineering development of such technologies. Nonetheless, only by inquiring into such concepts in a rigorous way can we hope to arrive at a proper assessment of the possibilities inherent in the evolution of advanced spaceflight technologies.

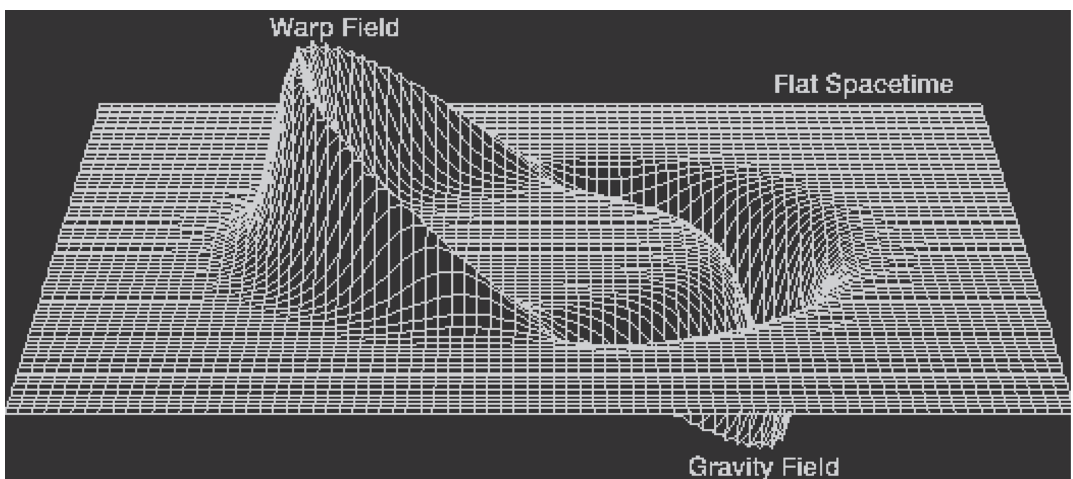

**Fig. 3 Alcubierre warp drive metric.**

## REFERENCES

1. M.G. Millis and E.W. Davis (eds), "*Frontiers of Propulsion Science*", AIAA Press, Reston, Virginia, 2009.
2. M. Alcubierre, "The Warp Drive: Hyper-fast Travel within General Relativity", *Class. Quantum Grav.*, **11**, pp.L73-L77, 1994.
3. H.E. Puthoff, "SETI, the Velocity-of-Light Limitation, and the Alcubierre Warp Drive: An Integrating Overview", *Physics Essays*, **9**, pp.156-158, 1996.
4. M.S. Morris and K.S. Thorne, "Wormholes in Spacetime and their Use for Interstellar Travel: A Tool for Teaching General Relativity", *Am. J. Phys.*, **56**, pp.395-412, 1988.
5. M. Visser, "*Lorentzian Wormholes: From Einstein to Hawking*", AIP Press, New York, 1995.
6. M.S. Morris, K.S. Thorne and U. Yurtsever, "Wormholes, Time Machines, and the Weak Energy Condition", *Phys. Rev. Lett.*, **61**, pp.1446-1449, 1988.
7. T.D. Lee, "*Particle Physics and Introduction to Field Theory*", Harwood Academic Press, London, 1988.
8. S. Saunders and H.R. Brown (eds), "*The Philosophy of Vacuum*", Clarendon Press, Oxford, 1991.
9. F. Wilczek, "*The Lightness of Being: Mass, Ether and the Unification of Forces*", Basic Books, New York, 2008.
10. A. Logunov and M. Mestvirishvili, "*The Relativistic Theory of Gravitation*", Mir Publ., Moscow, p.76, 1989.
11. Op. cit., p.83.
12. S.M. Mahajan, A. Qadir and P.M. Valanju, "Reintroducing the Concept of 'Force' into Relativity Theory", *Il Nuovo Cimento*, **65B**, pp.404-417, 1981.
13. R. Klauber, "Physical Components, Coordinate Components, and the Speed of Light", http://arxiv.org/abs/gr-qc/0105071. (Date Accessed 26 November 2010)
14. F. de Felice, "On the Gravitational Field Acting as an Optical Medium", *Gen. Rel. and Grav*, **2**, pp.347-357, 1971.
15. K.K. Nandi and A. Islam, "On the Optical-Mechanical Analogy in General Relativity", *Am. J. Phys.*, **63**, pp.251-256, 1995.
16. H.E. Puthoff, "Polarizable-Vacuum (PV) Approach to General Relativity", *Found. Phys.*, **32**, pp.927-943, 2002.
17. P. Boonserm *et al.*, "Effective Refractive Index Tensor for Weak-Field Gravity", *Class. Quant. Grav.*, **22**, pp.1905-1916, 2005.
18. X.-H. Ye and Q. Lin, "A Simple Optical Analysis of Gravitational Lensing", *J. Modern Optics*, **55**, pp.1119-1126, 2008.
19. H.E. Puthoff, E.W. Davis and C. Maccone, "Levi-Civita Effect in the Polarizable Vacuum (PV) Representation of General Relativity", *Gen. Relativ. Grav.*, **37**, pp.483-489, 2005.
20. A.P. Lightman and D.P. Lee, "Restricted Proof that the Weak Equivalence Principle Implies the Einstein Equivalence Principle", *Phys. Rev. D*, **8**, pp.364-376, 1973.
21. C.W. Misner, K.S. Thorne and J.A. Wheeler, "*Gravitation*", Freeman, San Francisco, p.5, 1973.
22. E.W. Davis, "Faster-than-Light Approaches in General Relativity", in *Frontiers of Propulsion Science*, Progress in Astronautics and Aeronautics Series, Vol. 227, eds. M.G. Millis and E. W. Davis, AIAA Press, Reston, VA, pp.471-507, 2009.
23. U. Leonhardt and T.G. Philbin, "General Relativity in Electrical Engineering", *New Jour. of Phys.*, **8**, pp.247-264, 2006.
24. J. Deardorff, B. Haisch, B. Maccabee and H.E. Puthoff, "Inflation Theory Implications for Extraterrestrial Visitation", *JBIS*, **58**, pp.43-50, 2005.

\* \* \*